\documentclass[a4paper,10pt,twoside]{cpc-hepnp}
\usepackage{amsmath, amssymb, graphicx, feynmp, adjustbox, ifthen}
\usepackage[colorlinks=true, citecolor=blue, urlcolor=blue, linkcolor=blue, breaklinks=true, pdfpagelabels=false]{hyperref}
\usepackage{subfigure}
\usepackage{float}
\usepackage{xcolor}
\usepackage{slashed}
\usepackage{multirow}
\usepackage{booktabs}
\usepackage{multicol}

\makeatletter\g@addto@macro\bfseries{\boldmath}\makeatother

\allowdisplaybreaks

\usepackage{CJKutf8}
\newcommand{\zw}[1]{\begin{CJK}{UTF8}{gbsn}{#1}\end{CJK}}

\newcommand{\FDF}{(\varphi^\dagger i\!\!\overleftrightarrow{D}_\mu\varphi)}
\newcommand{\FDFI}{(\varphi^\dagger i\!\!\overleftrightarrow{D}^I_\mu\varphi)}

\newcommand{\inab}{{\rm ab}^{-1}}
\newcommand{\infb}{{\rm fb}^{-1}}

\newcommand{\eehz}{e^+e^- \to hZ}
\newcommand{\eevvh}{e^+e^- \to \nu \bar{\nu} h}

\newcommand{\dkl}{\delta \kappa_\lambda}

\newcommand{\eett}{e^+e^- \to t\bar{t}}

\newcommand{\headertext}{Prepared for Chinese Physics C}
\newcommand{\preprintnumber}{}
\newlength{\preprintwidth}
\begin{document}

\newlength{\mycolwidth}
\setlength{\mycolwidth}{.5\textwidth}
\addtolength{\mycolwidth}{-.5\columnsep}

\begin{fmffile}{graph}
\setlength{\unitlength}{1mm}
\DeclareGraphicsRule{*}{mps}{*}{}
\fmfcmd{%
	prologues:=3;
	arrow_ang := 12;
	arrow_len := 3.5thick;
	color myg, myp, myr;
	myg := .7green;
	myp := .7blue;
	myr := .7red;
}
\newcommand{\myr}[1]{{\color{red!30!black}#1}}

\renewcommand{\preprintnumber}{\scriptsize{DESY 18-142, CERN-TH-2018-193, MITP/18-081}}

\title{Probing top-quark couplings indirectly at Higgs factories}

\author{Gauthier Durieux$^{1}$\email{gauthier.durieux@desy.de}%
\quad Jiayin Gu (\zw{顾嘉荫})$^{2}$\email{jiagu@uni-mainz.de}%
\quad Eleni Vryonidou$^{3}$\email{eleni.vryonidou@cern.ch}%
\quad Cen Zhang (\zw{张岑})$^{4}$\email{cenzhang@ihep.ac.cn}%
}

\maketitle

\address{%
$^1$ DESY Notkestraße 85, D-22607, Hamburg, Germany\\
$^2$ PRISMA Cluster of Excellence, Institut f\"ur Physik,\\
Johannes Gutenberg-Universit\"at, 55099 Mainz, Germany\\
$^3$ Theoretical Physics Department, CERN, 1211 Geneva 23, Switzerland\\
$^4$ Institute of High Energy Physics, and School of Physical Sciences,\\
University of Chinese Academy of Sciences, Beijing 100049, China\\
}
\begin{abstract}
	We perform a global effective-field-theory analysis to assess the combined
	precision on Higgs couplings, triple gauge-boson couplings, and
	top-quark couplings, at future circular $e^+e^-$ colliders, with a
	focus on runs below the $t\bar t$ production threshold. Deviations in the top-quark
	sector entering as one-loop corrections are consistently taken into
	account in Higgs and diboson processes. We find that future lepton
	colliders running at center-of-mass energies below the $t\bar t$ production threshold
	can still provide useful information on top-quark couplings, by
	measuring virtual top-quark effects.
        With rate and differential measurements, the indirect individual sensitivity achievable is better than at the high-luminosity LHC. However, strong correlations between the extracted top-quark and Higgs couplings are also present and lead to much weaker global
	constraints on top-quark couplings. This implies that a direct probe
	of top-quark couplings above the $t\bar t$ production threshold is helpful also for
	the determination of Higgs and triple-gauge-boson couplings.
	In addition, we find that below the $e^+e^-\to t\bar th$ production threshold, the top-quark Yukawa coupling can be
	determined by its loop corrections to all Higgs production and decay channels.
	Degeneracy with the $ggh$ coupling can be resolved,
	and even a global limit is competitive with the prospects of a linear collider
	above the threshold. This provides an additional means of determining
	the top-quark Yukawa coupling indirectly at lepton colliders.
\end{abstract}



\begin{multicols}{2}
\section{Introduction}
After the discovery of the Higgs boson~\cite{Aad:2012tfa, Chatrchyan:2012xdj}, 
understanding the electroweak symmetry breaking mechanism remains one of the major challenges in particle physics. The determination of Higgs
couplings at the Large Hadron Collider (LHC) is now approaching, and in some
cases surpassing, the 10\% precision level. Improvements beyond this level can
be foreseen at proposed $e^+e^-$ colliders. These
machines could run at a center-of-mass energy of 240--250~GeV---where the maximum of the $e^+e^-\to hZ$ cross section lies---or even above, and would
provide a much cleaner environment for precision determination of Higgs
couplings. Prospects have been widely studied through global analyses in the standard-model
effective field theory (SMEFT) and revealed that improvements of up to
several orders of magnitude can be achieved compared to present
limits~\cite{Ellis:2015sca, Ellis:2017kfi, Durieux:2017rsg, Barklow:2017suo,
Barklow:2017awn, DiVita:2017vrr, Chiu:2017yrx}.

Given the expected precision of measurements at future lepton colliders, next-to-leading-order
(NLO) theory predictions in the SMEFT can potentially be relevant.
These corrections can involve effective
operators which do not appear at leading order, and therefore provide new opportunities in the exploration of physics beyond the standard model (SM).
The indirect determination of the trilinear Higgs self-coupling, which enters single Higgs
production and decay processes at one loop, has for instance already been studied~\cite{McCullough:2013rea, DiVita:2017vrr, Chiu:2017yrx}.

Effective operators which give rise to anomalous $tbW$, $ttZ$,
$tt\gamma$ and $tth$ couplings of the top quark can also become relevant at one loop. If future lepton colliders run above the $e^+e^-\to t\bar t$ and $t\bar th$ production thresholds, these operator coefficients will be determined by
direct measurements (see Ref.~\cite{Durieux:2018tev} for a recent global study of
top-quark operators at future lepton colliders). Yet, at lower center-of-mass energies, they enter in loop corrections to other electroweak processes.
Recently, it has been pointed out that these
corrections are not negligible already at the LHC~\cite{Vryonidou:2018eyv}. Runs of future lepton colliders below the $t\bar{t}$ and $t\bar{t}h$ production thresholds may thus still provide complementary information on top-quark couplings.

Moreover, the top quark is an indispensable
player in Higgs coupling analyses due to its large Yukawa coupling. Already in the SM,
several important channels are dominated by its loop contributions.
Deviations which would be observed there could be sourced by the anomalous top-quark couplings.
However, it has been stressed in Ref.~\cite{Durieux:2017rsg} that NLO SMEFT predictions for processes that are not loop-induced in the SM are necessary for
a global and consistent fit at that order. Fortunately, computations at NLO in
the electroweak gauge couplings have become available for Higgs processes in the past few years~\cite{Hartmann:2015oia, Ghezzi:2015vva, Hartmann:2015aia, Gauld:2015lmb, Gauld:2016kuu, Dawson:2018pyl, Dedes:2018seb, Dawson:2018liq, Dawson:2018jlg}.
In particular, the NLO corrections involving top-quark operators for Higgs processes have become
available very recently~\cite{Vryonidou:2018eyv}, making such a combined
analysis feasible at future lepton colliders. The only missing ingredient was
the theory prediction for $W^+W^-$ production at the same order. This
process is notably sensitive to the triple gauge-boson couplings (TGC) which can be generated by operators affecting also Higgs interactions.

In this work, we extend the theory calculation and implementation of
Ref.~\cite{Vryonidou:2018eyv} with the one-loop contributions of top-quark operators to the
$e^+e^-\to W^+W^-$ process. It allows us to perform a consistent global fit in
the SMEFT, with Higgs couplings and TGCs at the tree level and
top-quark operators at the one-loop level. Our main focus is on future circular
lepton colliders with very good Higgs measurements but not large enough
center-of-mass energies to reach the $t\bar t$ or $t\bar th$ production thresholds.
We aim to answer the following questions:
\begin{itemize}
	\item If future lepton colliders only run below the $t\bar t$ ($t\bar th$) threshold, can they still determine top-quark--gauge-boson (top-quark Yukawa) couplings with high
		precision?
	\item Does the uncertainty on top-quark couplings affect
		the reach of future measurements of Higgs couplings?
\end{itemize}

The paper is organized as follows.
In Section~{\ref{sec:setup}}, we describe our theory framework.
In Section~{\ref{sec:calc}}, we review the calculation and implementation of
Ref.~\cite{Vryonidou:2018eyv}, and extend it to include $e^+e^-\to W^+W^-$ production.
In Section~{\ref{sec:fit}}, we describe the measurements and the fit.
We discuss our results in Section~{\ref{sec:res}} before concluding in Section~{\ref{sec:conclusion}}.
The likelihood of our fits are provided in an ancillary \href{https://arxiv.org/src/1809.03520/anc/chisquare-tloop.nb}{file} associated to our arXiv preprint.

\section{Effective-field-theory framework}
\label{sec:setup}
In the absence of clear signs of physics beyond the standard model (BSM), a common approach for testing the SM
and identifying possible deviations is provided by the SMEFT~\cite{Weinberg:1978kz, Leung:1984ni, Buchmuller:1985jz}. BSM effects are captured
by a series of higher dimensional operators whose coefficients can be related
to the parameters of specific models by a matching calculation. Given that all the
operators of odd dimension violate baryon or lepton numbers
\cite{Degrande:2012wf}, the most important deviations are expected to be captured by operators of dimension six:
\begin{flalign}
	\mathcal{L}_\text{EFT}=\mathcal{L}_\text{SM}+\sum_i\frac{C_i}{\Lambda^2}O_i^{(6)}
	+\dots  \label{eq:eft}
\end{flalign}
Measurements at the LHC and future lepton colliders can be conveniently interpreted in terms of their coefficients.

Two features of the SMEFT are of particular relevance to this work. The
first is that theory predictions can be improved systematically, order by
order. The SMEFT is a theory that is renormalizable order by order in
$1/\Lambda^2$~\cite{Weinberg:2009bg}. Thus, theory predictions can always be
improved to match experimental uncertainties. This is one of the main
advantages of the SMEFT over other BSM parametrizations, such as the anomalous coupling
approach to top-quark couplings, and the $\kappa$ framework for Higgs
couplings.

The second feature is that the SMEFT gives unambiguous and model-independent
results only if all operators up to a given dimension and up to a given loop order are simultaneously included.
This motivates the inclusion of top-quark operators at the one-loop level in all Higgs and diboson processes entering our global analysis and not only in loop-induced ones like $h\to gg$, $\gamma\gamma$ or $Z\gamma$.
In doing so, we include the contribution of each operator considered at its leading
order, i.e.\ at tree level for Higgs operators, and at one-loop level for most top-quark
operators. 
The tree- and loop-level contributions of other operators are not considered.
This may be justified either from a bottom-up or from a top-down point of view.
Without imposing restrictions on the type of BSM model covered by our SMEFT, one may argue that other operators are sufficiently constrained by measurements different from the ones considered here.
One may also argue that the class of models which would dominantly affect the top-quark and Higgs couplings through the operators we consider is worth studying.

Four-fermion operators giving rise to $e^+e^-t\bar t$ contact interactions are also disregarded although they could potentially play a role.
They contribute to Higgs and electroweak
processes once the top-quark line is closed in a loop. 
In particular, the two-fermion and the four-fermion
operators could not be efficiently discriminated if $\eett$ is only measured
near threshold~\cite{Durieux:2018tev}. So, without higher
energy runs, the Higgs and diboson measurements could potentially be used to
break this degeneracy.
Therefore, the inclusion of these operators could affect our results.
However, these corrections have not been computed so far and would affect the
renormalization of other SMEFT operators. The implementation of the
one-loop contributions of four-fermion operators as well as a full
analysis of their impact are therefore left to future study.
As these four-fermion operators are included in the global tree-level analysis
of Ref.~\cite{Durieux:2018tev}, we set their coefficients to zero when using
results from there. 
		
Our global analysis of Higgs and diboson measurements is based on that of Ref.~\cite{Durieux:2017rsg}.
Various observables are combined to constrain efficiently all directions
of the
multidimensional space spanned by the Higgs and top-quark operator
coefficients. They will be discussed in Section~{\ref{sec:fit}}.
We work under the same assumptions: departing from flavor universality only to
single out top-quark operators and distinguish the various measurable Yukawa
couplings, as well as taking electroweak and CP-violating
observables perfectly SM-like.\footnote{For studies concerning CP-violating top-Higgs interactions at future Higgs factories, see Refs.~\cite{Kobakhidze:2016mfx, LiuLowWang:2018}.}  
We also neglect the quadratic contributions
of dimension-six operators as justified in Ref.~\cite{Durieux:2017rsg}.
Operators that modify Higgs couplings and TGCs
are then captured by the following 12 parameters of the Higgs basis:
\begin{equation}
	\begin{gathered}
	\delta c_Z,\quad
	c_{ZZ},\quad
	c_{Z\square},\quad
	\bar c_{\gamma\gamma},\quad
	\bar c_{Z\gamma},\quad
	\bar c_{gg},
	\\
	\delta y_t,\quad
	\delta y_c,\quad
	\delta y_b,\quad
	\delta y_\tau,\quad
	\delta y_\mu,\quad
	\lambda_{Z}\,.
	\end{gathered}
\label{eq:higgs}
\end{equation}
As described in Ref.~\cite{Durieux:2017rsg} (with different notations), they can be easily mapped to the coefficients of 12
SILH-like basis operators:
\begin{equation}
\begin{aligned}
	&O_{\varphi W}
	=\varphi^\dagger \varphi W^I_{\mu\nu}W^{I\mu\nu}
	,\\
	&O_{\varphi \square}
	=\left( \varphi^\dagger\varphi \right)\square
	\left( \varphi^\dagger\varphi \right)
	,\\
	&O_{B}
	=iD^\mu\varphi^\dagger D^\nu\varphi B_{\mu\nu}
	,\\
	&O_{\mu\varphi}=(\varphi^\dagger\varphi)\bar l_2e_2\varphi+h.c.,
	\\
	&O_{t\varphi}=(\varphi^\dagger\varphi)\bar Qt\tilde\varphi+h.c.,
	\\
	&O_{WWW}=\epsilon^{IJK}W_\mu^{I\nu}W_\nu^{J\rho}W_\rho^{K\mu},
\end{aligned}
\quad
\begin{aligned}
	&O_{\varphi B}
	=\varphi^\dagger \varphi B_{\mu\nu}B^{\mu\nu}
	,\\
	&O_{W}
	=iD^\mu\varphi^\dagger\tau^ID^\nu\varphi W^I_{\mu\nu}
	,\\
	&O_{b\varphi}=(\varphi^\dagger\varphi)\bar Qb\varphi+h.c.
	,\\
	&O_{\tau\varphi}=(\varphi^\dagger\varphi)\bar l_3e_3\varphi+h.c.,
	\\
	&O_{c\varphi}=(\varphi^\dagger\varphi)\bar q_2u_2\tilde\varphi+h.c.,
	\\
	&O_{\varphi G}=\varphi^\dagger\varphi G_{\mu\nu}G^{\mu\nu},
\end{aligned}
\label{ops:2}
\end{equation}
where $Q$ is the third-generation quark doublet. The subscripts 2, 3 are flavor
indexes (weak and mass eigenstate fermions are not distinguished, approximating mixing matrixes by the identity). The assumption of perfect electroweak precision measurements in Ref.~\cite{Durieux:2017rsg} allowed to disregard the two operators
\begin{flalign}
	&O_{\varphi WB}
	=\varphi^\dagger\tau^I\varphi W^I_{\mu\nu}B^{\mu\nu}
	,
	&O_{\varphi D}
	=\left( \varphi^\dagger D^\mu\varphi \right)^*
	\left( \varphi^\dagger D_\mu \varphi \right)
	.
\end{flalign}
In this work, this assumption must be enforced at the one-loop level, including also top-quark operators. This will be discussed in the next section.

The 14 Higgs operators above form a set consistent with the basis employed in
the calculation of Ref.~\cite{Vryonidou:2018eyv}. The top-quark operators
considered here are the following:
\begin{align}\nonumber
	&O_{t\varphi}
	=\bar{Q} t\tilde\varphi\: (\varphi^{\dagger}\varphi)+h.c.
	,\\ \nonumber
	&O^{(1)}_{\varphi Q} =\FDF (\bar{Q}\gamma^\mu Q)
	,\\ \nonumber
	&O^{(3)}_{\varphi Q}
	=\FDFI (\bar{Q}\gamma^\mu\tau^I Q)
	,\\ \nonumber
	&O_{\varphi t}
	=\FDF (\bar{t}\gamma^\mu t)
	,\\ \nonumber
	&O_{tW} =(\bar{Q}\sigma^{\mu\nu}\tau^It)\:\tilde{\varphi}W_{\mu\nu}^I+h.c.
	,\\ \nonumber
	&O_{tB}
	=(\bar{Q}\sigma^{\mu\nu} t)\:\tilde{\varphi}B_{\mu\nu}+h.c.
	,\\ 
	&O_{tG}
	=(\bar{Q}\sigma^{\mu\nu} T^A t)\:\tilde{\varphi}G^A_{\mu\nu}+h.c.\,.
	\label{ops:1}
\end{align}
The $O_{\varphi tb}$ operator is neglected because its interferences with SM amplitudes are suppressed by a factor of $m_b$. In addition, we define
\begin{align}
	&O^{(+)}_{\varphi Q}\equiv
	\frac{1}{2}\left(  O^{(1)}_{\varphi Q}+O^{(3)}_{\varphi Q}\right),
	&O^{(-)}_{\varphi Q}\equiv
	\frac{1}{2}\left(  O^{(1)}_{\varphi Q}-O^{(3)}_{\varphi Q}\right),
\end{align}
and exclude $O_{\varphi Q}^{(+)}$ which affects the tightly constrained $Z\to
b\bar b$ branching fraction and asymmetry. Note that $O_{t\varphi}$ has been
included already in the Higgs operators, and its coefficient has a simple
relation with $\delta y_t$:\footnote{$\delta y_t$ receives an additional
	contribution from $C_{\varphi\square}$.  It is ommited because $\delta
	y_t$ in our calculation enters at the loop level, while we only aim at
	the LO contribution from $C_{\varphi\square}$.  }
\begin{equation}
	\delta y_t=-\frac{C_{t\varphi}v^2}{\Lambda^2}.
	\label{eq:yt}
\end{equation}
In summary, the following 6 top-quark operator coefficients are included in our analysis:
\begin{flalign}
	C_{\varphi t},\quad
	C_{\varphi Q}^{(-)},\quad
	C_{tW},\quad
	C_{tB}, \quad
	C_{t\varphi},\quad	
	C_{tG}. \label{eq:topop}
\end{flalign}

Apart from the top-quark operators, loop corrections also provide new
opportunities for indirectly constraining the Higgs trilinear coupling,
$\lambda_3$. The modification in the this
coupling is induced by a dimension-six operator $O_\varphi
=(\varphi^\dagger\varphi)^3$. The coupling can be directly constrained at the
LHC, but only at the $\mathcal{O}(1)$ level even assuming the high luminosity
scenario~\cite{ATL-PHYS-PUB-2017-001}. It was shown in
Ref.~\cite{McCullough:2013rea} that the measurements of the Higgsstrahlung
process at lepton colliders can have an indirect but competitive reach on this
coupling via its loop contribution. A global analysis was performed in
Ref.~\cite{DiVita:2017vrr}, which showed that the discrimination between the
Higgs trilinear coupling and other Higgs operators is possible, but nevertheless
nontrivial. In this work, to determine the impact of $\lambda_3$ on the
global reach of the top-quark operators, we follow Ref.~\cite{DiVita:2017vrr} and
include its one-loop contribution to all the single Higgs processes,
parameterized by $\delta \kappa_\lambda \equiv \kappa_\lambda -1$, where
$\kappa_\lambda$ is the ratio of the Higgs trilinear coupling to its SM value,
\begin{equation}
\kappa_\lambda \equiv \frac{\lambda_3}{\lambda_3^{\rm SM}}\,, \qquad \lambda_3^{\rm SM} = \frac{m_h^2}{2 v^2}\,.
\end{equation}
By turning on and off this coupling in our fit, we will see by how much
the determination of top-quark couplings will be affected.


\section{Theory predictions}
\label{sec:calc}
To the precision needed for this work, the theory predictions for, e.g., total cross sections can
be written as
\begin{equation}
\begin{aligned}
	\sigma= \sigma_{\rm SM} 
	&+ C_h(\mu_\text{EFT})\:\sigma_\mathrm{tree}\\
	&+ C_t(\mu_\text{EFT})\frac{\alpha_{EW}}{\pi}
	\!\! \left( \sigma_\mathrm{log} \log\frac{Q^2}{\mu_\text{EFT}^2} +\sigma_\mathrm{fin}  \!\! \right)
	.
\end{aligned}
\end{equation}
Here, $C_h(\mu_\text{EFT})$ is the coefficient of some Higgs or TGC operator $O_h$ that
contributes at the tree level, and $\mu_\text{EFT}$ is the scale at which the
coefficient is defined. In this work we take $\mu_\text{EFT}=m_H$ for all
measurements.
$C_t$ is the coefficient of some top-quark operator $O_t$
which enters at the loop level and could potentially mix into $O_h$. $Q^2$ is the
scale of the process. The calculation of $\sigma_\mathrm{tree}$ is
straightforward while $\sigma_\mathrm{log}$ can be obtained from the running
of SMEFT coefficients. In this section, we review the computation of the genuine electroweak corrections $\sigma_\mathrm{fin}$ carried out in Ref.~\cite{Vryonidou:2018eyv} for Higgs processes.
We then compute them for $e^+e^-\to W^+W^-$ production.

\begin{center}
\vspace{5mm}
\begin{fmfgraph*}(25,15)
\fmfstraight
\fmfleft{l1,l2}
\fmfright{r1,r2}
\fmf{fermion,foreground=myg}{l2,v1,l1}
\fmfdot{v2}
\fmf{photon,foreground=myp,tens=3}{v1,v2}
\fmf{fermion,foreground=myr,label=$\myr{t}$, l.side=left}{v2,w2}
\fmf{photon,foreground=myp}{w2,r2}
\fmf{fermion,foreground=myr}{w1,v2}
\fmf{photon,foreground=myp}{r1,w1}
\fmffreeze
\fmf{fermion,foreground=myr}{w2,w1}
\end{fmfgraph*}
\quad
\begin{fmfgraph*}(25,15)
\fmfstraight
\fmfleft{l1,l2}
\fmfright{r1,r2}
\fmf{fermion,foreground=myg}{l2,v1,l1}
\fmfdot{v3}
\fmf{photon,foreground=myp,tens=3}{v1,v2}
\fmf{photon,foreground=myp}{v3,r2}
\fmf{photon,foreground=myp}{r1,v3}
\fmf{phantom, tens=2}{v3,v2}
\fmffreeze
\fmf{fermion,left,foreground=myr,label=$\myr{t}$}{v2,v3}
\fmf{fermion,left,foreground=myr}{v3,v2}
\end{fmfgraph*}
\\[5mm]
\begin{fmfgraph*}(25,15)
\fmfstraight
\fmfleft{l1,l2}
\fmfright{r1,r2}
\fmf{fermion,foreground=myg}{l2,v1,l1}
\fmf{photon,foreground=myp,tens=3}{v1,v2}
\fmf{phantom, tens=2}{v2,v3}
\fmfdot{v3}
\fmf{photon,foreground=myp,tens=3}{v3,v4}
\fmf{photon,foreground=myp}{v4,r2}
\fmf{photon,foreground=myp}{r1,v4}
\fmffreeze
\fmf{fermion,left,foreground=myr,label=$\myr{t}$}{v2,v3}
\fmf{fermion,left,foreground=myr}{v3,v2}
\end{fmfgraph*}
\quad
\begin{fmfgraph*}(25,15)
\fmfstraight
\fmfleft{l1,l2}
\fmfright{r1,r2}
\fmf{fermion,foreground=myg}{l2,w2,w1,l1}
\fmf{phantom}{w1,r1}
\fmf{phantom}{w2,r2}
\fmffreeze
\fmf{photon,foreground=myp,tens=3}{w2,v2}
\fmf{fermion,left,foreground=myr,label=$\myr{t}$}{v2,v3}
\fmf{fermion,left,foreground=myr}{v3,v2}
\fmfdot{v3}
\fmf{photon,foreground=myp,tens=3}{v3,r2}
\fmf{photon,foreground=myp}{w1,r1}
\end{fmfgraph*}
\vspace{5mm}
\figcaption{Selected diagrams for dimension-six top-quark contributions to $e^+e^-\to W^+W^-$.
Red lines represent the top quark. Blobs represent dimension-six operator insertions.
\label{fig:eeww}}
\end{center}

The complete set of electroweak NLO corrections from top-quark operators to precision
electroweak operators was first given in Ref.~\cite{Zhang:2012cd}. Results can
conveniently be obtained in the ``star scheme''~\cite{Peskin:1991sw}, because all
contributions are oblique. For Higgs production this is not any longer the case. 
In addition to the $VV$ self-energy corrections one has to compute also $hh$
and $hVV$ functions, where $V$ is a photon, $W$ or $Z$ boson. While several
calculations were available in the literature~\cite{Hartmann:2015oia,
Ghezzi:2015vva, Hartmann:2015aia, Gauld:2015lmb, Dawson:2018pyl,
Dedes:2018seb, Dawson:2018liq}, 
the complete results for top-quark operator contributions to Higgs
production in the $Vh$ and VBF channels, as well as decay modes $h\to
\gamma\gamma,\gamma Z,Wl\nu,Zll,b\bar b,\mu\mu,\tau\tau$ were first presented
in Ref.~\cite{Vryonidou:2018eyv}.
This excludes the four-fermion operators mentioned previously.
The calculation is implemented in the
MadGraph5\_aMC@NLO framework~\cite{Alwall:2014hca} whose reweighting
functionality~\cite{Mattelaer:2016gcx} is used to compute the dimension-six top- and bottom-quark loop
contributions. The SM parameters are renormalized consistently in the $m_W$,
$m_Z$ and $G_F$ scheme up to dimension six, and operator coefficients are renormalized
in the $\overline{MS}$ scheme. The rational R2 counterterms are computed following the scheme
of Ref.~\cite{Kreimer:1989ke, Korner:1991sx, Kreimer:1993bh}, for $ZZ$, $hh$, $hVV$,
$ffV$ and $ffh$ loop functions. The implementation provides an automatic and
convenient way to simulate indirect contributions from top-quark operators, which
enter Higgs processes as NLO electroweak corrections. Events
can be generated and matched to parton shower, allowing for detailed
investigations using the full differential information.

It is well known that, in the SMEFT formalism, the measurements of Higgs
couplings and TGCs are entangled~\cite{Pomarol:2013zra,Corbett:2013pja,Falkowski:2015jaa}. $W$ pair
production is therefore an important component of global Higgs
analyses at future lepton colliders.
For this reason, we extend the calculation of Ref.~\cite{Vryonidou:2018eyv} to
incorporate the $e^+e^-\to W^+W^-$ process. Some diagrams involving dimension-six
operators are shown in Fig.~\ref{fig:eeww}. Additional counterterms need to be computed for the
$WW\gamma$ and $WWZ$ vertexes. Among the three TGC operators,
only $O_W$ and $O_B$ are renormalized by top-quark operators. The
anomalous dimensions are derived in Ref.~\cite{Vryonidou:2018eyv}. Another
difficulty is that the $WW\gamma$ function involves a triangle anomaly diagram. 
In our
scheme, this implies that the R2 counterterms depend on the
choice of the vertex from which the trace of the
fermion loop starts. This effect is in principle canceled by a Wess-Zumino
term generated when chiral fermions in the full theory are integrated out~\cite{Wess:1971yu}. The problem can be fixed by imposing the Ward identity of the
photon in the low-energy effective theory. We provide more details in
Appendix~\ref{app:anomaly}. We have validated our implementation of the
$WW\gamma$ vertex by computing processes with an external photon and checking
that the Ward identity is satisfied.

Our global analysis relies on the assumption that precision electroweak measurements
are perfectly constrained to be SM-like. This has consequences on our renormalization scheme, as explained in the following.

In our operator basis, precision electroweak observables receive tree-level contributions from $O_{\varphi WB}$ and $O_{\varphi D}$ operators.
At that order, their coefficients are thus simply removed from the fit by assuming the measurements of precision electroweak observables perfectly match SM predictions.
Top-quark operators however start contributing at the loop
level. In the $\overline{MS}$ scheme,
the same assumption implies that $C_{\varphi WB}$ and $C_{\varphi D}$ need to
take specific values to cancel these loop corrections. These nonzero values
will then in turn modify other Higgs production and decay channels, making the fit
more complicated. In Ref.~\cite{Vryonidou:2018eyv}, a more
convenient approach has been followed, where $C_{\varphi WB}$ and $C_{\varphi D}$ are defined in the
on-shell scheme using oblique parameters as renormalization conditions.
Therefore, if the oblique parameters are tightly constrained, we can exclude
$C_{\varphi WB}$ and $C_{\varphi D}$ from the fit.

Instead of using oblique parameters, we further refine this approach in this
work, by using the full set of $Z$-pole and $W$-pole measurements listed in
Ref.~\cite{Patrignani:2016xqp} as our renormalization conditions. 
We assume that, apart from deviations in the top-quark
and the Yukawa sectors, BSM effects are otherwise universal, in the sense that
they can be captured by dimension-six operators which involve SM bosons only,
up to suitable field redefinitions~\cite{Wells:2015uba}. Their effects in
Higgs, $WW$, and $Z/W$-pole measuremetns are then fully captured by operators
listed in Section~{\ref{sec:setup}}.\footnote{In the SILH basis, two additional
operators $O_{2B}=-\frac{1}{2}\left( \partial^\mu B_{\mu\nu}\right)^2$
and $O_{2W}=-\frac{1}{2}\left( D^\mu W_{\mu\nu}^a\right)^2$ are
universal, but they can be eliminated in favor of four-fermion operators, and
thus drop out from the pole measurements.} With this assumption, the two most
constraining degrees of freedom from $Z$-pole and $W$-pole measurements can
then be used as physical observables to renormalize the coefficients of
$O_{\varphi WB}$ and $O_{\varphi D}$ operators. In the following, we briefly
describe the procedure.

According to Ref.~\cite{Zhang:2012cd}, contributions from dimension-six
top-quark operators to precision electroweak measurements can be conveniently evaluated in the $\alpha$, $m_Z$ and $G_F$ scheme,
by taking the SM tree-level predictions at the $Z$ pole, written in terms of
$\alpha$, $m_Z$ and $s_W$, and making the following substitutions:%
\end{multicols}%
\vspace{2mm}
\ruleup[\mycolwidth]
\begin{eqnarray}%
\alpha&\rightarrow&\alpha_*=\alpha+\delta\alpha
=\alpha\left(1-\Pi'_{\gamma\gamma}(q^2)+\Pi'_{\gamma\gamma}(0)\right)
\times\left[
1-\frac{d}{dq^2}\Pi_{ZZ}(q^2)|_{q^2=m_Z^2}+\Pi'_{\gamma\gamma}(q^2)+\frac{c_W^2-s_W^2}{s_Wc_W}\Pi'_{\gamma Z}(q^2)
\right],
\label{replace3}\\
m_Z^2&\rightarrow&m_{Z*}^2=m_Z^2+\delta m_Z^2
=m_Z^2+\Pi_{ZZ}(m_Z^2)-\Pi_{ZZ}(q^2)+(q^2-m_Z^2)\frac{d}{dq^2}\Pi_{ZZ}(q^2)|_{q^2=m_Z^2},\label{replace35}
\\
s_{W}^2&\rightarrow&s_{W*}^2=s_{W}^2+\delta s_{W}^2
=s_W^2\left[
1+\frac{c_W}{s_W}\Pi'_{\gamma Z}(q^2)+\frac{c_W^2}{c_W^2-s_W^2}\left(
\Pi'_{\gamma\gamma}(0)+\frac{1}{m_W^2}\Pi_{WW}(0)-\frac{1}{m_Z^2}\Pi_{ZZ}(m_Z^2)
\right)
\right].\label{replace4}
\end{eqnarray}%
\ruledown[\mycolwidth]
\vspace{5mm}
\begin{multicols}{2}%
\noindent and taking $q^2=m_Z^2$. Here $\Pi_{VV}(q^2)$ is the self-energy
correction for the $V=\gamma,W,Z$ gauge boson while $\Pi'_{VV}(q^2)\equiv
\left[\Pi_{VV}(q^2)-\Pi_{VV}(0)\right]/q^2$. Expressions for these corrections
are of order $C/\Lambda^2$ and can be found in Ref.~\cite{Zhang:2012cd}.
Note that unlike in the calculation for Higgs and $WW$ production,
here we use $G_F$ instead of $m_W$ as an input parameter, since theory
predicitons in Ref.~\cite{Patrignani:2016xqp} are provided in this scheme.

$Z$-pole observables consist of various combinations of partial widths and asymmetries.
Their SM predictions only depend on $s_W$ and on the product of $\alpha$ and $m_Z$. Thus
these measurements only constrain two independent combinations of top-quark
operator coefficients. The correction to the $W$ mass can be written as
\begin{equation}%
	\begin{aligned}
	\frac{\delta m_W^2}{m_W^2}&=
	\frac{s_W^2}{c_W^2-s_W^2}\Pi'_{\gamma\gamma}(0)
	+\frac{c_W^2}{c_W^2-s_W^2}\frac{1}{m_W^2}\Pi_{WW}(0)
	\\&-\frac{c_W^2}{c_W^2-s_W^2}\frac{1}{m_Z^2}\Pi_{ZZ}(m_Z^2)
	+\Pi'_{WW}(m_W^2)\,.
	\end{aligned}
\end{equation}%
and constitutes the third independent combination constrained. Finally, the
width of the $W$ boson is corrected by
\begin{multline}
	\delta\left( \frac{\Gamma_W}{m_W} \right)^2=
	\left( \frac{\Gamma_W}{m_W} \right)^2\times
	\\
	\left( \frac{\delta m_W^2}{m_W^2} -\Pi'_{WW}(m_W^2)
	+\frac{d}{dq^2}\Pi_{WW}(q^2)|_{q^2=m_W^2}\right)
\end{multline}
which is almost degenerate with the previous constraint. Given the relatively weaker
precision on the measurement of $\Gamma_W$ and the approximate degeneracy, we
expect this fourth constraint to be less useful.

We now modify the renormalization of $C_{\varphi WB}$ and
$C_{\varphi D}$ by finite $\Delta_{ij}$ constants, so that
\begin{flalign}
	&C_i\;\Rightarrow\; Z_{ij}C_j=C_i+\delta Z_{ij} C_j,\\
	&\delta Z_{ij}=\frac{\alpha}{2\pi}\Gamma(1+\epsilon)
	\left( \frac{4\pi \mu^2}{\mu_\text{EFT}^2}\right)^\epsilon 
	\left(\frac{1}{\epsilon}
	+\Delta_{ij}\right)\gamma_{ij}
\end{flalign}
for $O_i$ = $O_{\varphi WB}$, $O_{\varphi D}$. The $O_j$ cover all top-quark
operators. We then need to choose the values of $\Delta_{ij}$ which minimize the deviations in the precision observables when setting $C_{\varphi WB}=C_{\varphi D}=0$. To find them, we
construct a $\chi^2$ using the experimental results and theory predictions
for all $W$ and $Z$ pole data listed in Ref.~\cite{Patrignani:2016xqp}. The
associated covariance matrix has four positive eigenvalues, corresponding to the four independent
constraints expected. The $\Delta_{ij}$ are chosen such that the two most
constrained eigenvectors only involve $C_{\varphi WB}$ and $C_{\varphi D}$:
\begin{flalign}
	&+0.906 C_{\varphi WB} + 0.423 C_{\varphi D} = 0\pm 0.0000234
	\\
	&-0.423 C_{\varphi WB} + 0.906 C_{\varphi D} = 0\pm 0.0124
\end{flalign}
for $\Lambda=1$~TeV. In this specific scheme and up to one-loop order,
the two most stringent limits from $Z$- and $W$-pole data only constrain the renormalized
$C_{\varphi WB}$ and $C_{\varphi D}$ to small values. The assumption of
perfect precision measurements approximates these two limits to be infinitely constraining and allows us to exclude $C_{\varphi WB}$ and $C_{\varphi D}$ from the rest of our analysis at the one-loop level.
This can be interpreted as using the first two degrees of freedom of the
precision measurements as on-shell renormalization conditions for these two
coefficients.

There are two remaining constraints.
One is associated with a covariance matrix eigenvalue of about 300~TeV$^{-2}$, is the
weakest. We therefore ignore it. The other implies
$0.17 C_{\varphi Q}^{(-)} - 0.10 C_{\varphi t}  - 0.04 C_{tB} - 0.92 C_{tW} = 0
\pm 0.60 \times (\Lambda/\mathrm{TeV})^2$ 
and involves only top-quark operators. We include this constraint in our fit,
conservatively assuming that it could be strengthened by a factor of five with future
lepton collider data.
Our final results are however largely insensitive to this constraint.

The above renormalization scheme as well as the $W$-pair process are then supplemented to the UFO model described in Ref.~\cite{Vryonidou:2018eyv}. It allows for the automatic calculations of all Higgs and $W$-pair processes
relevant to this work. Beside inclusive cross sections,
differential distributions can also be obtained. The production angle in
$e^+e^-\to W^+W^-$ will be used in our analysis.

Our discussion so far excluded the top-quark chromo-dipole operator
$O_{tG}$. It enters $h\to gg$ through a top-quark loop which has already been
studied in the literature~\cite{Degrande:2012gr, Maltoni:2016yxb}. This
effect will be included in our global analysis.

\section{Measurements and fit}
\label{sec:fit}

In this section, we describe the measurements and observables used in our
analysis. Since our study is most relevant for lepton colliders with very good
Higgs measurements but not large enough center-of-mass energies to reach the
$t\bar{t}$ or $t\bar{t}h$ production thresholds, our primary focus will be on the circular
colliders. Currently, two proposals for such colliders have been made: the
Circular Electron Positron Collider (CEPC) in China~\cite{CEPC-SPPCStudyGroup:2015csa}, and the Future Circular Collider with
$e^+e^-$ beams (FCC-ee) at CERN~\cite{Gomez-Ceballos:2013zzn}.
In this study, we consider the following hypothetical scenario:
a circular collider (CC) collecting an integrated luminosity of $5~\inab$ at a center-of-mass energy of 240~GeV and possibly also running at the top-quark pair production threshold with $0.2~\inab$ gathered at 350~GeV
and $1.5~\inab$ at 365~GeV. The incoming electron and positron beams are
assumed to be unpolarized. This scenario follows closely the current projected
run plan of the FCC-ee~\cite{Benedikt:2018}. The 350 and 365~GeV runs could fix the
top-quark electroweak couplings by directly probing $t\bar t$ production, though
approximate degeneracies would remain due to limited energy lever arm. The top-quark Yukawa coupling, on the other hand, cannot be directly probed in such a run scenario.
We will also show results with only a 240~GeV run, which represents the
CEPC scenario. At the moment, there is no plan for the CEPC
to run at center-of-mass energies beyond 240~GeV, though a future upgrade to
the top-quark pair production threshold remains an open possibility. Our study could then
provide useful information regarding the impact of a 350~GeV upgrade on the
measurements of the top-quark couplings and on the indirect effect of their loop contributions on Higgs coupling determinations.
Both CEPC and FCC-ee plan to also run
at the $Z$-pole and $WW$ production threshold, which could significantly improve the
sensitivities on the electroweak observables that are already tightly
constrained by LEP measurements.

Linear colliders, such as the  Compact Linear Collider at CERN~\cite{CLIC:2016zwp} and the International Linear Collider~\cite{Baer:2013cma},
could run at higher energies and, in particular, above the thresholds for both $t\bar{t}$ and
$t\bar{t}h$ productions. With these measurements, the top-quark operators can be
probed at the tree level, with a sensitivity far better than the current one.
Although information about top-quark couplings may only be indirectly available at the first 250~GeV stage of the ILC, we will not treat this case explicitly and will not further consider linear collider scenarios.

\begin{figure*}[t]
\centering
\includegraphics[width=0.7\textwidth]{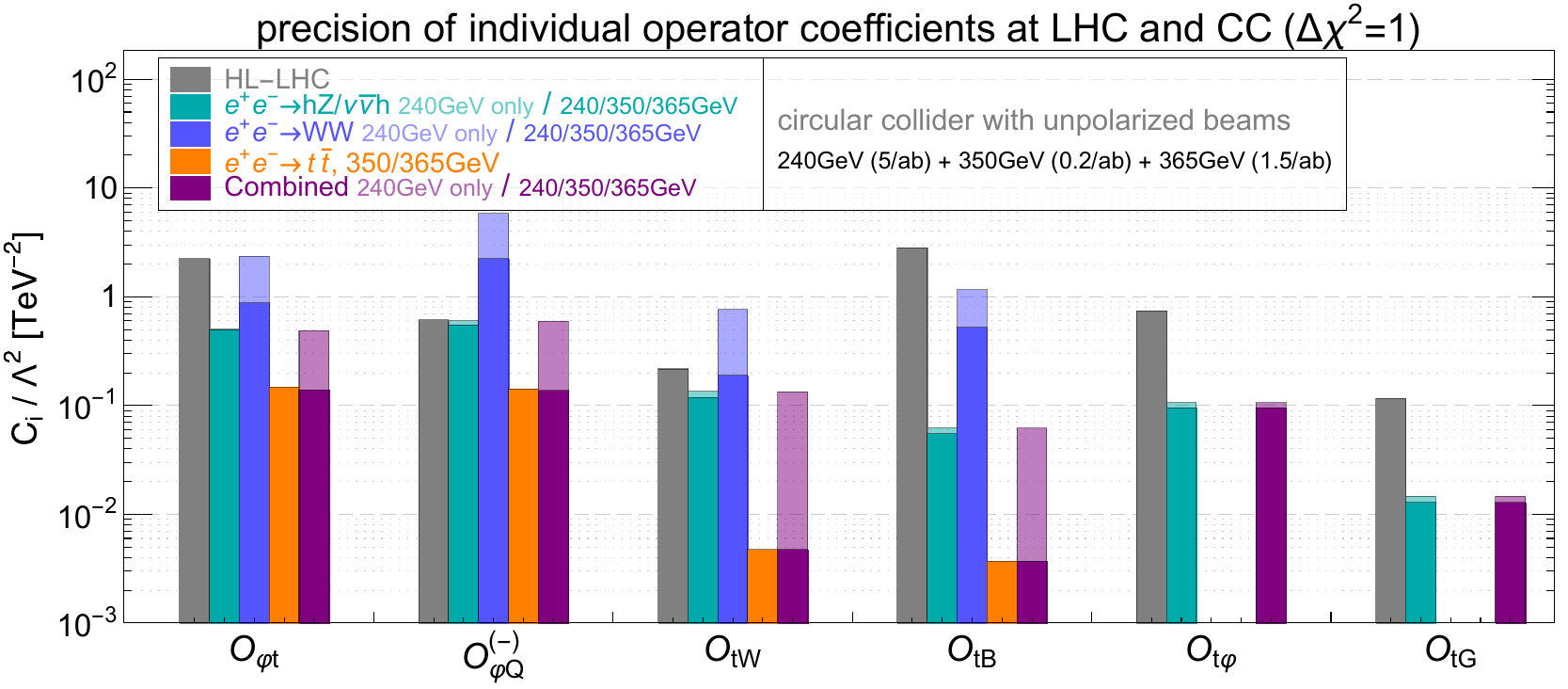}
\caption{\label{fig:barccindv1}Individual one-sigma reach on top-quark operator coefficients for different future collider scenarios and measurements.
One single coefficient is allowed to depart from zero at a time.
}
\end{figure*}

\begin{center}
\tabcaption{\label{tab:hl-lhc-measurements}Estimates for the precision reachable on key top-quark observables at the HL-LHC.}
\begin{tabular*}{\columnwidth}{@{\extracolsep{\fill}}|c|cc|}
	\hline
	Channels & \multicolumn{2}{c|}{Uncertainties}
	\\
	& without th.\ unc. & with th.\ unc.
	\\\hline
	$t\bar t$ & 4\% \cite{Sirunyan:2017uhy} & 7\%
	\\
	Single top ($t$-ch.) & 4\% \cite{Schoenrock:2013jka} & 4\%
	\\
	$W$-helicity ($F_0$) & 3\% \cite{Aaboud:2016hsq} & 3\%
	\\
	$W$-helicity ($F_L$) & 5\% \cite{Aaboud:2016hsq} & 5\%
	\\
	$t\bar{t}Z$ & 10\% & 15\%
	\\
	$t\bar{t}\gamma$ & 10\% & 17\%
	\\
	$t\bar{t}h$ & 10\% & 16\% \cite{ATL-PHYS-PUB-2014-016}
	\\
	$gg\to h$ & 4\% & 11\%  \cite{ATL-PHYS-PUB-2014-016}
	\\\hline
\end{tabular*}%
\end{center}

The HL-LHC measurements could provide important complementary information. 
To the best of our knowledge, a systematic determination of the projected sensitivity to top-quark couplings
is unfortunately not available in the literature.
We therefore consider the measurements in Table~\ref{tab:hl-lhc-measurements} and estimate the precision reachable with the HL-LHC.
Here, the projected precisions on measurements of the $t\bar t$ and $t$-channel single top-quark production cross sections and $W$-helicity in top-quark decay are based on the works referred to. 
Theoretical uncertainties for single top-quark production and $W$-helicity measurements are neglected,
as they are both of order $\mathcal{O}(1\%)$~\cite{Berger:2016oht, Czarnecki:2010gb}.
The uncertainties on $t\bar th$ and $gg\to h$ production cross sections are taken from Ref.~\cite{ATL-PHYS-PUB-2014-016}.
The ones imposed on the $t\bar tZ/\gamma$ cross sections are simple estimates.
Theoretical uncertainties are estimated from predictions at NLO in QCD.
A combination of $t\bar t$, $t\bar th$ and $gg\to h$ production cross section measurements is
sufficient to constrain $O_{tG}$, $O_{t\varphi}$ and $O_{\varphi G}$ ($\bar
c_{gg}$). 
The $W$-helicity measurements alone fix $O_{tW}$. The remaining
three operators, $O_{\varphi t}$, $O_{\varphi Q}^{(-)}$ and $O_{tB}$ are
constrained by $t\bar tZ/\gamma$ and single top-quark production cross section measurements.
For the trilinear Higgs self-coupling, we follow Ref.~\cite{DiVita:2017eyz, Azatov:2015oxa} and assume that a constraint of $-0.9<\dkl<1.3$ at the $\Delta\chi^2=1$ level could be obtained at the HL-LHC by measuring both the rate and the distributions of the double Higgs production process. While a global fit should in principle be performed also for the HL-LHC, it was shown in Ref.~\cite{DiVita:2017eyz} that the reach on the trilinear Higgs coupling is dominated by the measurement of the double Higgs production, while the other Higgs operators are well constrained by the single Higgs processes and have little impact on the extraction of the trilinear Higgs coupling. We expect this to hold even with the inclusion of top-quark operator contributions in the loops.
The combination of measurements in Table~\ref{tab:hl-lhc-measurements}
with that of double Higgs production captures the most important
information on
top-quark operators and the trilinear Higgs self-coupling at the HL-LHC.

For the Higgs measurements at lepton colliders, we follow closely the
treatment of Ref.~\cite{Durieux:2017rsg} and include both the inclusive $\eehz$
cross section and exclusive Higgs decay channels,
as well as the measurement of the $WW$-fusion production channel, $\eevvh$.
The run scenario in Ref.~\cite{Durieux:2017rsg} has been updated to the one
detailed above. While the differential observables in $\eehz$ could provide
additional information~\cite{Beneke:2014sba, Craig:2015wwr}, they are not
included in our analysis. For these observables, corrections in production and
decay of the Higgs and $Z$ need to be simulated simultaneously, and this is not
yet possible in our setup.  For the diboson production process, $e^+e^-\to
W^+W^-$, we consider only the semileptonic decay channel, assuming the
statistical uncertainties dominate.  In contrast with
Ref.~\cite{Durieux:2017rsg}, we only include the differential distributions of
the $W$-production polar angle.
Finally, for the measurements of $t\bar{t}$ production at center-of-mass
energies of 350 and 365~GeV, we use the results of Ref.~\cite{Durieux:2018tev}.
We do not consider the one-loop corrections to $\eett$ from the top-quark
operators, since most of them enter at tree level and can therefore be tightly
constrained.  In particular, the loop-level dependence in the top-quark Yukawa
and chromo-dipole operators are not accounted for.
The total $\chi^2$ is obtained by summing over the $\chi^2$ of all the
measurements whose central values are assumed to confirm SM predictions.

It was shown in Ref.~\cite{Durieux:2017rsg} that, thanks to the high precision
of the measurements at the lepton colliders, it is sufficient to only keep the
linear dependences of the observables on the EFT parameters. We found this
statement to hold even with the inclusion of the top-quark operator in loops.
However, at the HL-LHC, the cross section for $t\bar t$ production in
association with a $Z$ boson or photon has a limited sensitivity to the linear
contributions of $O_{tW}$ and $O_{tB}$~\cite{Bylund:2016phk}, due to the
Lorentz structure of these dipole operators and to accidental cancellations
between different initial states. The inclusion of the $W$-helicity
measurement significantly improves the reach on $O_{tW}$ and brings its
dependence in the fit back to the linear regime. For $O_{tB}$, on the other
hand, the reach is much worse and is mainly driven by the quadratic terms. The
inclusion of these terms would significantly complicate our fitting procedure.
Since our focus is rather on the lepton colliders, 
we simplify the fit by keeping only the linear terms while adding by hand an extra term to the total $\chi^2$ that corresponds to a standard deviation of 3~TeV$^{-2}$ for $C_{tB}/\Lambda^2$, which reproduces the main constraints on $O_{tB}$ from the square contributions to a good approximation.

\begin{figure*}\centering
\includegraphics[width=0.45\textwidth]{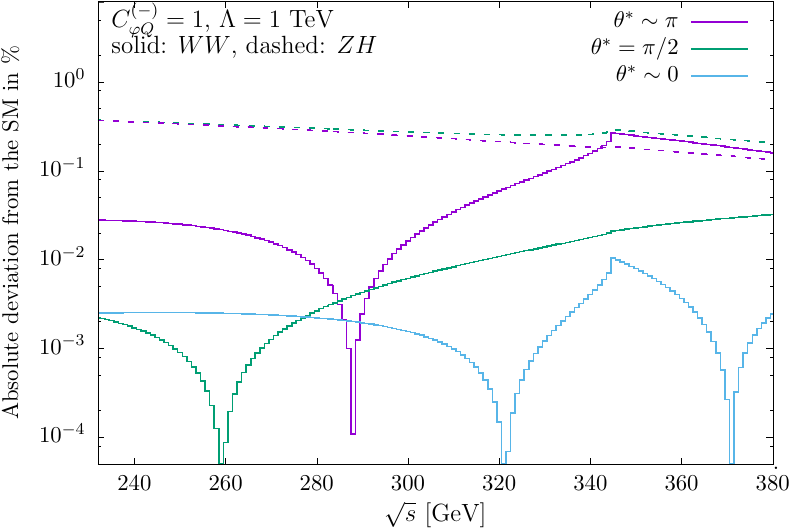}
\qquad
\includegraphics[width=0.45\textwidth]{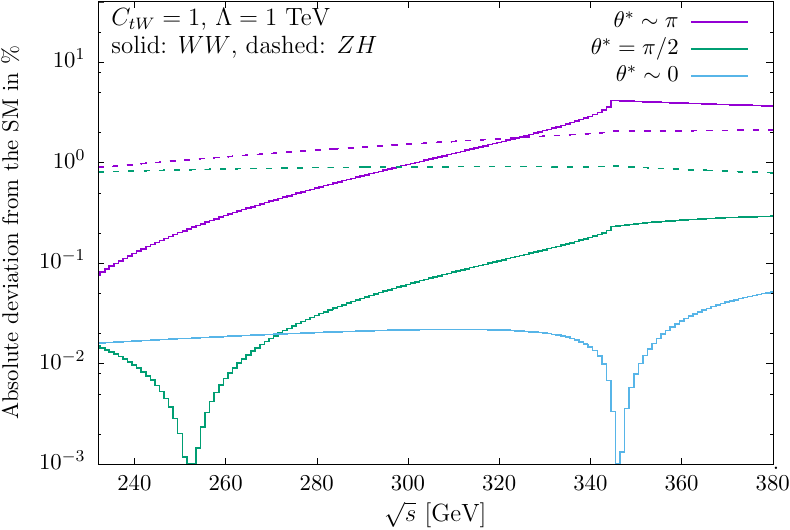}
\caption{\label{fig:wwenergy}Center-of-mass energy dependence of $O_{\varphi Q}^{(-)}$ (left) and $O_{tW}$ (right) contributions to $e^+e^-\to
W^+W^-$ (solid) and $Zh$ (dashed) production in percent of the SM rate, for $C/\Lambda^2 = 1$~TeV$^{-2}$.
Dependences are shown in absolute values (changes of sign generate the visible dips) for three different scattering angles: 0, $\pi/2$ and $\pi$.
For $e^+e^-\to Zh$, the $0$ and $\pi$ curves overlap.
}
\end{figure*}

\section{Results}
\label{sec:res}

In this section, we present the precision reach obtained by a $\chi^2$ fit to the observables described in
the previous section.
There are two important aspects in the determination of the
indirect reach on the top-quark operators from the Higgs and diboson measurements at lepton colliders.
First, the overall measurement sensitivity is assessed by performing individual fits to each parameter of
Eq.~(\ref{eq:topop}), setting all others to zero.
The second aspect concerns
the discrimination among top-quark operators, and between the Higgs and top-quark ones.
Differential information is then crucial to constrain all directions of the SMEFT parameter space with a limited number of processes.
Runs at different center-of-mass energies and with different polarizations also help setting meaningful constraints with lepton collider data only, though the latter information is not available at circular colliders.
One can otherwise also resort to a combination with HL-LHC measurements.

The individual sensitivities to the top-quark operators of Eq.~(\ref{eq:topop}) are shown in Fig.~\ref{fig:barccindv1} for different measurements at a circular lepton collider as well as at the HL-LHC. The results are presented in terms of the one-sigma reach
on $C_i/\Lambda^2$, with $C_i$ and $\Lambda$ defined in Eq.~(\ref{eq:eft}). Five scenarios are considered. The first column corresponds to the HL-LHC measurements listed in Table~\ref{tab:hl-lhc-measurements}, with theoretical uncertainties included.
The second, third and forth columns respectively include the Higgs, diboson, and $t\bar{t}$ measurements at a circular lepton collider.
The last column is obtained from the combination of all these circular collider measurements.
Lighter shades are obtained with a 240~GeV run only, while the darker ones combine operation all three center-of-mass energies considered (240, 350 and 365~GeV).

The indirect individual reach of Higgs and diboson measurements at 240~GeV on top-quark operator coefficients is seen to be better than the direct HL-LHC sensitivity.
The loop suppression of top-quark operator contributions is compensated by the high precision of lepton colliders measurements.
This is one of the main conclusions of this work, and partly answers the first
question raised in the introduction.
If higher center-of-mass energies are available, direct $e^+e^-\to t\bar{t}$ measurements still provide the best reach on top-quark operators.
Note that the tree-level analysis of the top-quark pair production from Ref.~\cite{Durieux:2018tev} is insensitive to the top-quark Yukawa and chromo-dipole operators.
The indirect individual reach of the diboson measurements at 240~GeV is somewhat lower than that of Higgs measurements.
It however improves with 350/365~GeV runs. This higher indirect sensitivity of $W$ pair production to top-quark operators at higher center-of-mass energies is further examined in Fig.~\ref{fig:wwenergy}.
For illustration, the contributions of $O_{\varphi Q}^{(-)}$ and $O_{tW}$ operators are shown in percent of the SM rate, as a function of the center-of-mass energy and for three scattering angles.
For comparison, dashed curves show the dependence of $e^+e^-\to Zh$ production.

Among top-quark operators, the improvement brought by 240~GeV Higgs
measurements over HL-LHC individual sensitivities is most significant for
$O_{tB}$. It mainly arises from the measurement of the Higgs decay to two
photons, $h\to \gamma\gamma$. As shown in Ref.~\cite{Vryonidou:2018eyv}, the
$O_{tB}$ operator has a particularly large contribution to this decay channel:
roughly of the same size as the SM rate when
$C_{tB}/\Lambda^2=(1~\mathrm{TeV})^{-2}$. On the other hand, no direct
measurements at the LHC can probe $O_{tB}$ efficiently. It should be noted
that the measurements of $h\to \gamma\gamma$ at the HL-LHC could also help probing
$O_{tB}$. Similarly, $O_{t\varphi}$ and $O_{tG}$ can be
very well constrained individually by the measurement of $h\to gg$ at lepton colliders. However,
we will see in our global analysis that the degeneracy
between the top-quark and Higgs operators is to be partially lifted by loop
corrections in other Higgs processes.

\begin{figure*}[t]
\centering
\adjustbox{max width=\textwidth}{%
\includegraphics{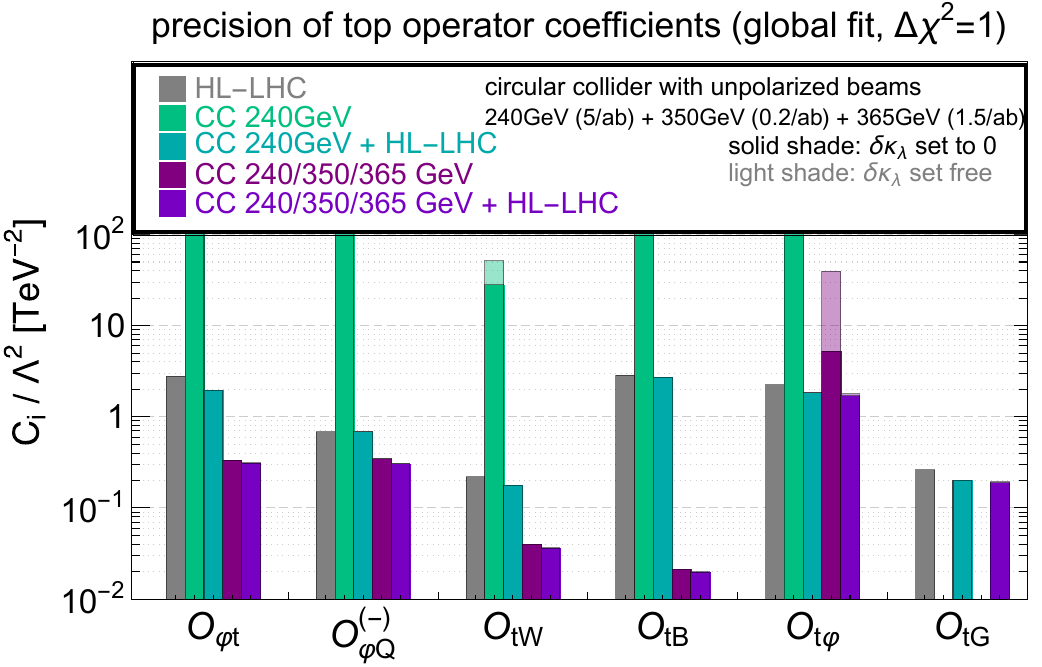}%
\includegraphics{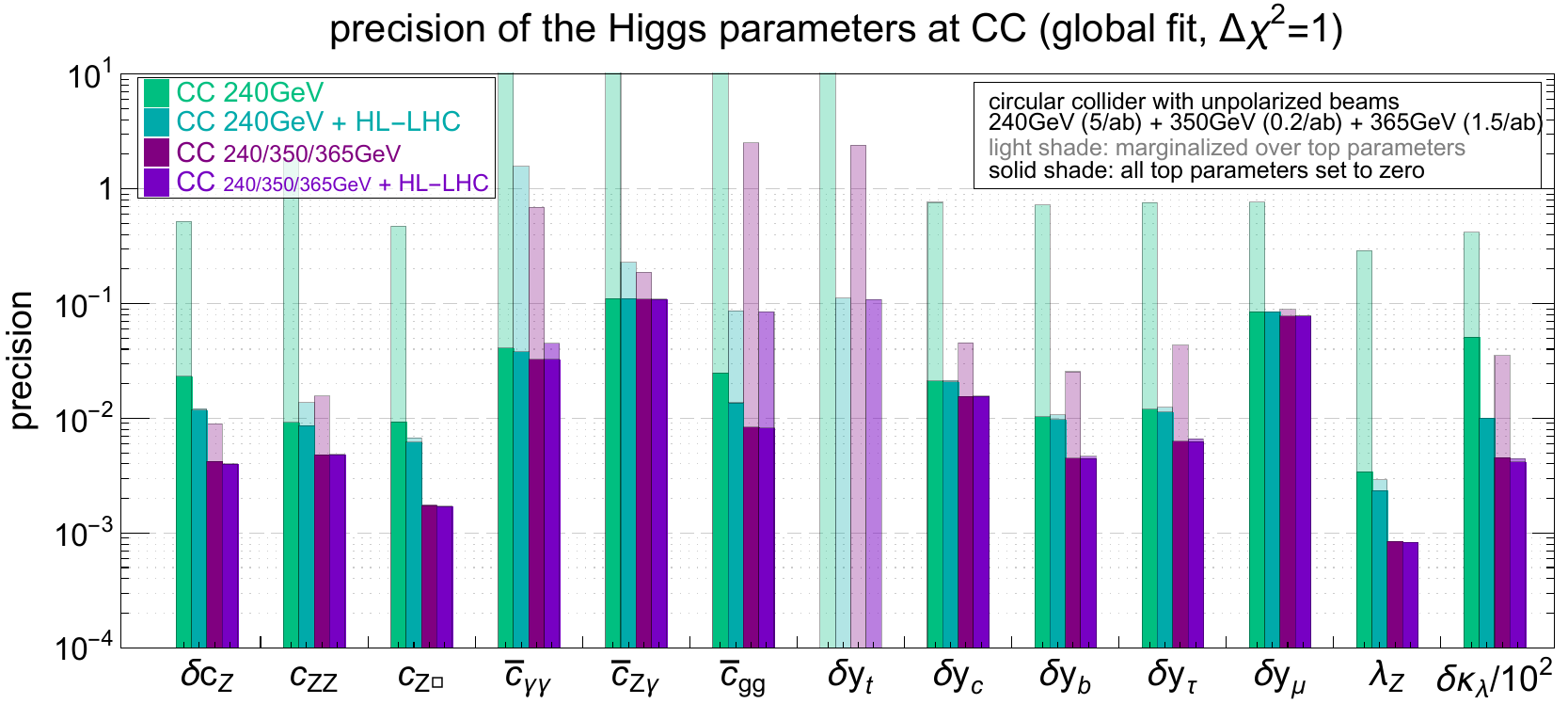}}
\caption{\label{fig:barccglobal1}Global one-sigma precision reach on the 18
top-quark (left) and Higgs (right) operator coefficients deriving from HL-LHC
and circular lepton collider measurements. The Higgs parameter definitions are
that of Ref.~\cite{Durieux:2017rsg}.  Large degeneracies are present in the CC
240\,GeV scenario and push the precision reach on some operator coefficients
outside of the plot range.  With lepton-collider measurements only, $C_{tG}$
and $\bar{c}_{gg}$ remain fully correlated. The constraint displayed for
$\bar{c}_{gg}$ is then actually to be interpreted as applying on $\bar{c}_{gg}+
0.46 \, C_{tG}$.
}
\end{figure*}

In Fig.~\ref{fig:barccglobal1}, we present the results of the global analysis
for all Higgs and top-quark operator coefficients of Eq.~(\ref{eq:higgs}) and
(\ref{eq:topop}). It amounts to 18 degrees of freedom once the trilinear Higgs
boson self-coupling is included. Note that $\delta y_t$ and $C_{t\varphi}$
represent the same degree of freedom since they are related through
Eq.~(\ref{eq:yt}). The reach on the top-quark and Higgs operator coefficients is
respectively shown in the left and right panels. For top-quark operators, five
scenarios are presented. The first column shows the reach of the HL-LHC
measurements.
The second column shows the indirect reach of a 240~GeV run. This result
is then combined with the HL-LHC measurements and displayed in the third column.
The fourth and the fifth columns display similar information, but with all
three energies, 240, 350 and 365~GeV. The $\eett$ measurements are then in particular included. We also
display the impact of $\dkl$ on the reach of the top-quark operators. The
results shown with the light shades are obtained by setting $\dkl$ to zero, and
the ones with darker shades are obtained by marginalizing over $\dkl$.
The impact of $\delta\kappa_\lambda$ is small once the double Higgs measurements
of the HL-LHC are included.

As expected, the indirect global reach of Higgs and diboson measurements on
top-quark operator coefficients is much lower than the individual one.  In
particular, large degeneracies are present when data from a 240~GeV run only
is exploited, pushing global limits beyond the range of validity of the EFT. While the
dependence of observables used in the fit on dimension-six operator coefficients
is still dominated by linear contributions, these limits should be interpreted
with care.
The difference between individual and global constraints is particularly
pronounced for $C_{tB}$, $C_{t\varphi}$ and $C_{tG}$ due to their approximate degeneracies with Higgs operators.
The $h\to \gamma\gamma$ branching fraction is for instance very well constrained but, alone, does not  discriminate between the contributions from $C_{t\varphi}$, $C_{tB}$ and $\bar c_{\gamma\gamma}$.
Similarly, $h\to gg$ measurements only constrain a combination of $C_{t\varphi}$,
$C_{tG}$ and $\bar c_{gg}$.
Lepton collider runs nevertheless provide some marginal improvement in a combination with direct top-quark measurements at the HL-LHC.
Note that the $O_{tG}$ operator enters $h\to gg$ but no other measurement
at 240~GeV. So its marginalized limit without combination with HL-LHC
data is absent. 
At higher energies, it could enter in NLO corrections to $t\bar t$ production (or in $t\bar tj$) which we do not include.
This is in contrast with $O_{t\varphi}$ whose marginalized limit
at lepton colliders derive from its loop corrections to other
channels which are however not loop-induced.
We will further discuss the reach on the top-quark Yukawa coupling at the end of this section.
Direct measurements of $e^+e^-\to t\bar{t}$ still yield the best handle on top-quark operator coefficients.
As mentioned earlier, it remains to be examined whether they are also efficient in constraining indirectly the $O_{t\varphi}$ and $O_{tG}$ operator coefficients in a global analysis.
In our treatment, the main constraints on these parameters arise from the HL-LHC measurements of $t\bar t$, $t\bar{t}h$, and $gg\to h$.

In the right panel of Fig.~\ref{fig:barccglobal1}, the one-sigma reach on Higgs couplings are presented for circular lepton colliders with and without combination with HL-LHC data.
The impact of a 240~GeV run alone is again separated from that of the full scenario considered, with operation at center-of-mass energies of 240, 350 and 365~GeV.
In this figure, we aim to answer the second question raised in the introduction,
by emphasizing the impact of uncertainties on top-quark couplings on the extraction
of Higgs couplings.
This is visible in the difference between bars of lighter and darker shades,
for which the corresponding top-quark operator coefficients (including $\delta y_t$) are respectively
marginalized over or set to zero.
Considering a lepton collider run at 240~GeV only,
without any direct constraint on top-quark operator coefficients, these
uncertainties typically worsen the reach on most Higgs couplings by more than
one order of magnitude. Several global limits ---on $\bar c_{\gamma\gamma}$,
$\bar{c}_{Z\gamma}$ and $\bar c_{gg}$ in particular--- are then too loose to
remain meaningful.
The impact of the top-quark loop contributions on most Higgs couplings is
significantly reduced once direct top-quark pair production measurements are performed above the $\eett$ production threshold.
The uncertainty on the top-quark Yukawa coupling still sizeably affects the determination of several Higgs boson couplings.
The HL-LHC data cures this issue for all couplings but $\bar c_{gg}$.
Without lepton collider run above the $t\bar t$ production threshold, the loose constraint on $C_{tB}$ deriving from HL-LHC measurements degrades the global limit on $\bar c_{\gamma\gamma}$ by more than one order of magnitude.

\begin{center}
\includegraphics[width=0.35\textwidth]{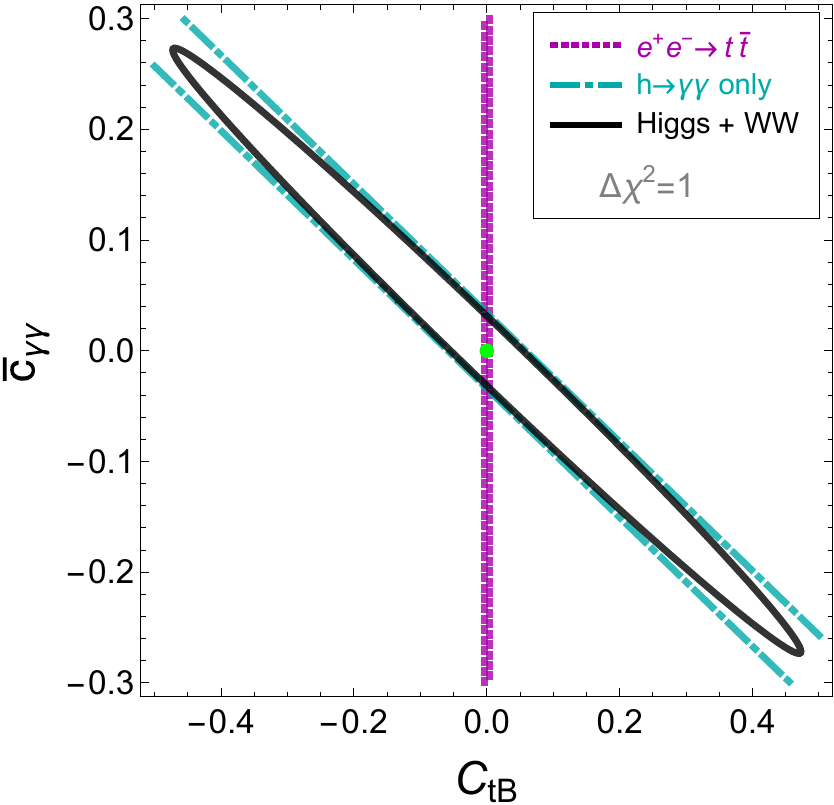}
\figcaption{\label{fig:tbaa}Two-dimensional constraints on $C_{tB}$ and $\bar c_{\gamma\gamma}$, with
all other parameters set to zero, to illustrate the correlation between Higgs and top-quark couplings.}
\end{center}

This correlation is further examined in Fig.~\ref{fig:tbaa} showing the individual $\Delta\chi^2=1$ sensitivities of various measurements in the two-dimensional parameter space formed by $C_{tB}$ and $\bar{c}_{\gamma\gamma}$.
The $h\to\gamma\gamma$
measurement imposes a tight constraint on a linear combination of $C_{tB}$ and
$\bar c_{\gamma\gamma}$, leading to a strong correlation between these two
parameters, but also leaving a blind direction unconstrained.
The latter can be lifted either at lepton collider via loop corrections involving $O_{tB}$ to other processes, or at the HL-LHC via direct $ttZ/\gamma$
measurements, but none of them is strong enough to simultaneously pin down both couplings.
In particular, HL-LHC measurements yield a loose $-2.7<C_{tB}<2.1$ constraint
for $\Lambda=1$~TeV which cannot be displayed in Fig.~\ref{fig:tbaa}.
As already stressed, direct $\eett$ measurements above 350~GeV are needed to resolve this issue.
Similar observations can also be made for $\bar c_{Z\gamma}$. The lower precision achieved on the $hZ\gamma$ interaction somewhat reduces the impact of correlations in that case.

\begin{center}
\includegraphics[width=0.48\textwidth]{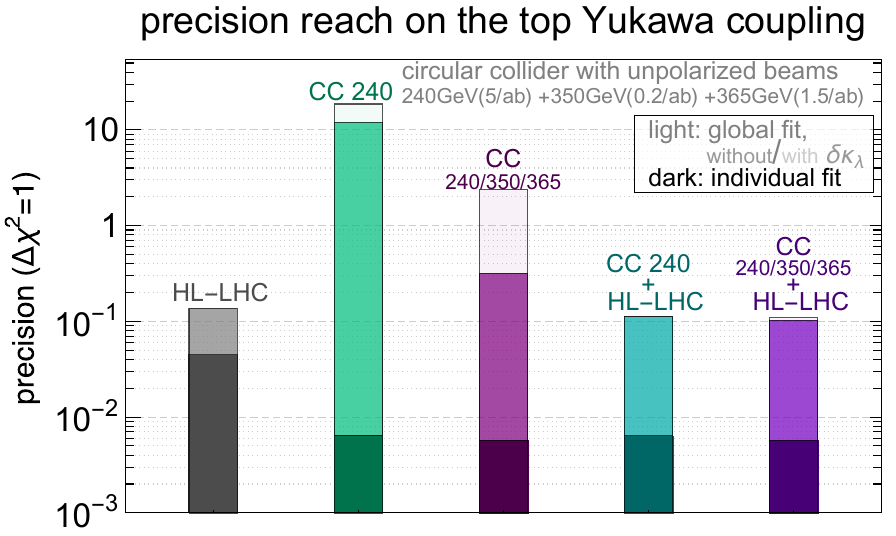}
\figcaption{\label{fig:yt}Indirect one-sigma reach on $\delta y_t$ in different lepton collider scenarios, compared and combined with the HL-LHC measurements.}
\end{center}

Finally, we show in Fig.~\ref{fig:yt} the indirect reach on the top-quark
Yukawa coupling, $\delta y_t$, from Higgs and diboson measurements at a
circular lepton collider. With only a 240~GeV run at a circular lepton
collider, a strong correlation with $\bar c_{gg}$ makes the global reach on
$\delta y_t$ about three orders of magnitude weaker than the individual one.
The individual reach is dominated by the precision of the $h\to gg$ branching
fraction measurement (see also Ref.~\cite{Boselli:2018zxr}).
In contrast, the global one is determined by loop-level
sensitivity of processes that are not loop-induced. Additional runs at
center-of-mass energies of 350 and 365~GeV directly fix top-quark--gauge-boson
couplings through $\eett$ measurements and improve the global constraint on
$\delta y_t$ by more than an order of magnitude. Still, an approximate
degeneracy with the loop-dependence on the trilinear Higgs self-coupling is
visible and is only resolved by a combination with HL-LHC
measurements.\footnote{Conversely, the impact of the uncertainty on $\delta y_t$ in the extraction of $\dkl$ through loop corrections in $e^+e^-\to hZ$ at 240~GeV was studied in Ref.~\cite{Shen:2015pha}.} The loop-level
sensitivity of $\eett$ on $\delta y_t$ which we did not include
is potentially complementary. With a CLIC beam
spectrum (broader than that of a circular collider), a $t\bar{t}$
threshold scan alone leads to a precision of about $20\%$ on $\delta
y_t$ determined simultaneously with the top-quark mass using a total
integrated luminosity of $100~\infb$~\cite{Abramowicz:2018rjq}.
Setting $\dkl$ to zero, the indirect sensitivity of Higgs and diboson
processes in runs at center-of-mass energies of 240, 350 and 365~GeV
leads to a global one-sigma precision of $32\%$ on $\delta y_t$. This reach is
competitive with the one achievable at the HL-LHC. To compare with
direct measurements, the $e^-e^+\to t\bar th$ production cross section
with $1~\inab$ of integrated luminosity collected at a center-of-mass
energy of 500~GeV with a $P(e^+,e^-)=(+0.3,-0.8)$ beam polarization
would lead to a precision of $10\%$ on $\delta y_t$~\cite{Yonamine:2011jg}.
We thus conclude that the loop contributions to Higgs and diboson processes
studied in this work provide an additional handle on $\delta y_t$ below the
$t\bar th$ threshold, leading a global reach competitive with that of other
direct and/or indirect approaches. This completes the answer to the first
question in our introduction.

\section{Conclusions}
\label{sec:conclusion}
In this work, we have studied the sensitivity of a future circular $e^+e^-$
collider to Higgs couplings, triple gauge-boson couplings, and top-quark couplings. In
particular, we focused on runs below the $e^+e^-\to t\bar t$ production threshold, where
top-quark couplings enter as one-loop corrections. The corrections to the
Higgs processes became available in Ref.~\cite{Vryonidou:2018eyv}. We have
obtained the corrections to $W$-boson pair production which were not previously
known. Based on these results, we have performed a global SMEFT analysis
including both Higgs and $W$-pair measurements.
This allowed us to derive the future sensitivities to all couplings considered simultaneously.

The main finding of this work is that future lepton colliders running at
center-of-mass energies below the $t\bar t$ threshold can provide useful
information on top-quark couplings through the measurements of virtual effects.
The indirect individual sensitivities obtained are higher than the direct HL-LHC
ones.
Nevertheless, our analysis suggests that an energy upgrade above the $e^+e^-\to t\bar t$ production threshold is desirable.
On the one hand, the direct individual sensitivity to top-quark couplings is much higher.
On the other hand, the strong correlations between the top-quark and Higgs couplings which manifest themselves in a global analysis are mitigated.
Below the $t\bar t$ threshold, global constraints on top-quark couplings are otherwise much weaker than individual ones, if meaningful at all.  
The combination a $240$~GeV run with direct top-quark coupling measurements at the HL-LHC does not entirely solve this issue.
A precise determination of top-quark couplings is thus also crucial for fixing Higgs couplings.

In addition, we find that lepton colliders running below the $t\bar th$ production
threshold can also determine the top-quark Yukawa coupling through its loop
corrections to other Higgs channels. Combining 240~GeV and
350/365~GeV runs leads to a marginalized limit that is
competitive with projected direct limits at the HL-LHC as well as at the ILC with 500~GeV of center-of-mass energy.
Higgs and diboson measurements thus provide an alternative indirect determination of the top-quark Yukawa coupling at future circular lepton collider,
beside a $t\bar t$ threshold scan. Given that latter is also affected
by the mass of the top quark and the former by loops of the trilinear Higgs self-coupling, the two approach are expected to be complementary.
This interplay should be further studied in the future.
Note that the 350/365~GeV runs are crucial for the precision
of this approach. This provides another motivation for the corresponding energy
upgrade at circular lepton colliders.

A few simplifications have been made throughout our analysis. Four-fermion
and CP-odd operators were not included, as the corresponding electroweak NLO corrections are yet not available.
Top-quark pair production at lepton colliders was treated at tree level.
Precision electroweak measurements were assumed to be infinitely constraining.
Our approach could be applied to the lower-energy stages of a linear collider where beam polarization would provide an additional handle.
A more extensive use of differential distributions could also improve the reach we presented here and help lifting approximate degeneracies.
Further investigations along these directions can be envisioned.

\section*{Acknowledgements}
We thank C.~Grojean and M.~Riembau for helpful discussions about the fit,
X.~Zhao for useful discussions about renormalization schemes, and
Y.~Bai and K.~Mimasu for useful discussions about chiral anomaly in the SMEFT.
CZ is supported by IHEP under Contract No.~Y7515540U1. EV is supported
by a Marie Sk\l{}odowska-Curie Individual Fellowship of the European 
Commission's Horizon 2020 Programme under contract number 704187.

\appendix
\section{Gauge anomaly in the \texorpdfstring{$WW\gamma$}{WWA} vertex}
\label{app:anomaly}

Effective operators could induce gauge anomalies by modifying the top-quark
couplings to gauge bosons, which are chiral. In our scheme, this is reflected
by the fact that the R2 rational counterterms of the $W^+W^-\gamma$ loop function contain
 a term with the epsilon tensor, whose coefficient depends on
the vertex from which we compute the fermionic trace.
In the following, we list the epsilon term in the R2 counterterms for all
relevant operators, with the fermion loop traced from all three vertexes,
$\gamma$, $W^+$, and $W^-$. Our convention is that the three external fields,
$A^\mu$, $W^{+\nu}$, and $W^{-\rho}$, are associated with incoming momenta
$p_1$, $p_2$ and $p_3$ respectively. They are:
\begin{flalign}
&	O_{\varphi Q}^{(+)}:
	\quad
	-\frac{e^3v^2}{48\pi^2s_W^2\Lambda^2}
	\left\{
                \begin{array}{ll}
			\epsilon^{\mu\nu\rho\sigma}(p_{2\sigma}-p_{3\sigma}) &\quad \gamma
			\\
			\epsilon^{\mu\nu\rho\sigma}(p_{3\sigma}-p_{1\sigma}) &\quad W^+
			\\
			\epsilon^{\mu\nu\rho\sigma}(p_{1\sigma}-p_{2\sigma}) &\quad W^-
                \end{array}
              \right.
	      \label{eq:r21}
\\
&	O_{\varphi Q}^{(-)}:
	\quad
	\frac{e^3v^2}{48\pi^2s_W^2\Lambda^2}
	\left\{
                \begin{array}{ll}
			\epsilon^{\mu\nu\rho\sigma}(p_{2\sigma}-p_{3\sigma}) &\quad \gamma
			\\
			\epsilon^{\mu\nu\rho\sigma}(p_{3\sigma}-p_{1\sigma}) &\quad W^+
			\\
			\epsilon^{\mu\nu\rho\sigma}(p_{1\sigma}-p_{2\sigma}) &\quad W^-
                \end{array}
              \right.
\\
&	O_{tB}:
	\quad
	\frac{3e^2c_Wvm_t}{8\sqrt{2}\pi^2s_W^2\Lambda^2}
	\left\{
                \begin{array}{ll}
			0 &\quad \gamma
			\\
			\epsilon^{\mu\nu\rho\sigma}p_{1\sigma} &\quad W^+
			\\
			-\epsilon^{\mu\nu\rho\sigma}p_{1\sigma} &\quad W^-
                \end{array}
              \right.
\\
&	O_{tW}:
	\quad
	\frac{e^2vm_t}{8\sqrt{2}\pi^2s_W\Lambda^2}
	\left\{
                \begin{array}{ll}
			-3\epsilon^{\mu\nu\rho\sigma}(p_{2\sigma}-p_{3\sigma}) &\quad \gamma
			\\
			2\epsilon^{\mu\nu\rho\sigma}(p_{1\sigma}-p_{2\sigma}) &\quad W^+
			\\
			-2\epsilon^{\mu\nu\rho\sigma}(p_{1\sigma}-p_{3\sigma}) &\quad W^-
                \end{array}
              \right.
\end{flalign}
The field after each line indicates the starting point of the trace.
The other operators do not contribute.

This anomaly can be interpreted as the consequence of integrating out heavy chiral fermions.
The anomaly free condition in the UV theory implies anomaly cancellation between different fermions.
When matching to the SMEFT, if only some of them are integrated out, the resulting effective field
theory could appear to be anomalous. However, when these chiral fermions are integrated out, they
also generate a Wess-Zumino term which is supposed to cancel the gauge anomaly in the
SMEFT. This term has the following from:
\begin{flalign}
	c_{WZ}\frac{e^3}{8\pi^2s_W^2}\epsilon^{\mu\nu\rho\sigma}A_\mu\left
	(W_\nu^I\partial_\rho W^I_\sigma+\frac{1}{3}g_W \epsilon_{IJK}W_\nu^I W_\rho^J W_\sigma^K\right)
\end{flalign}

The coefficient of this term can be determined by requiring that the Ward identity for $U(1)_\text{EM}$ is
restored in the effective theory. Taking $O_{\varphi Q}^{(+)}$ as an example, we first go to the consistent anomaly~\cite{Bardeen:1984pm} by symmetrizing
the anomaly with respect to all three external momenta. From Eq.~(\ref{eq:r21}), this corresponds to a
vanishing R2 counterterm, and
\begin{flalign}
	p_1^\mu \Gamma_{\mu\nu\rho}=p_2^\mu\Gamma_{\rho\mu\nu}=p_3^\mu\Gamma_{\nu\rho\mu}=
	-\frac{C_{\varphi Q}^{(+)}e^3v^2}{48\pi^2s_W^2\Lambda^2}\epsilon^{\nu\rho\alpha\beta}p_{2\alpha}p_{3\beta}
	\label{eq:consistentanomaly}
\end{flalign}
Then, the Wess-Zumino term gives an additional contribution
\begin{flalign}
	&\Gamma_{\mu\nu\rho}^{WZ}=c_{WZ}\frac{e^3}{8\pi^2s_W^2}\epsilon^{\mu\nu\rho\sigma}(p_{2\sigma}-p_{3\sigma})
\\
&p_1^\mu\Gamma_{\mu\nu\rho}^{WZ}=2c_{WZ}\frac{e^3}{8\pi^2s_W^2}\epsilon^{\mu\nu\alpha\beta}p_{2\alpha}p_{3\beta}
\end{flalign}
For this to cancel the anomaly in Eq.~(\ref{eq:consistentanomaly}), we need
\begin{flalign}
	c_{WZ}=\frac{C_{\varphi Q}^{(+)}v^2}{12\Lambda^2}
\end{flalign}
In our implementation, the contribution from this term can be added together with the R2 counterterms, leading to
\begin{flalign}
	\mathrm{R2_{O_{\varphi Q}^{(+)}}}(WW\gamma)=\frac{C_{\varphi Q}^{(+)}e^3v^2}{96\pi^2s_W^2\Lambda^2}\epsilon^{\mu\nu\rho\sigma}(p_{2\sigma}-p_{3\sigma})
\end{flalign}
All three other operators can be dealt with in the same way. In practice, we note
that this is equivalent to computing the trace by starting from $W^+$ and $W^-$
respectively, and then taking the average. Finally, we use the same
prescription for the $WWZ$ vertex.

\vspace{3mm}\nopagebreak[4]
\bibliographystyle{apsrev4-1_title}
\bibliography{tloop}

\end{multicols}
\end{fmffile}
\end{document}